\documentclass[aps, pre, floatfix, superscriptaddress, twocolumn, footinbib, sort&compress, numbers, merge, reprint]{revtex4}

\usepackage{graphicx}
\usepackage{amsmath}
\usepackage{amssymb}
\usepackage{booktabs}
\usepackage{color}
\usepackage{natbib}
\usepackage{chngcntr}
\usepackage[utf8]{inputenc}
\usepackage[tight,sf]{subfigure}

\usepackage{hhline}
\usepackage[colorlinks=true, linkcolor=blue, citecolor=blue, filecolor=blue, urlcolor=blue]{hyperref}
\usepackage[all]{hypcap}

\newcommand{\Tensor}[1]{\mathbf{#1}}
\newcommand{\UnitVec}[1]{\mathbf{\hat{#1}}}

\begin{document}

\title{Geometric control of bacterial surface accumulation}

\author{Rachel Mok}
\affiliation{Department of Mechanical Engineering, Massachusetts Institute of Technology, 77 Massachusetts Avenue, Cambridge,~MA~02139-4307, USA}
\affiliation{Department of Mathematics, Massachusetts Institute of Technology, 77 Massachusetts Avenue, Cambridge,~MA~02139-4307, USA}
\author{J{\"{o}}rn Dunkel}
\affiliation{Department of Mathematics, Massachusetts Institute of Technology, 77 Massachusetts Avenue, Cambridge,~MA~02139-4307, USA}
\author{Vasily Kantsler}
\affiliation{Department of Physics, University of Warwick, Coventry CV4 7AL, United Kingdom}

\date{\today}
   
\begin{abstract}
Controlling and suppressing bacterial accumulation at solid surfaces is essential for preventing biofilm formation and biofouling. Whereas various chemical surface treatments are known to reduce cell accumulation and attachment, the role of complex surface geometries remains less well understood. Here, we report experiments and simulations that explore the effects of locally varying boundary curvature on the scattering and accumulation dynamics of swimming \textit{Escherichia coli} bacteria in quasi-two-dimensional microfluidic channels. Our experimental and numerical results show that a concave periodic boundary geometry can decrease the average cell concentration at the boundary by more than 50\% relative to a flat surface.
\end{abstract}
 
\pacs{}

\maketitle

\section{Introduction}
In the vicinity of surfaces, the behavior of swimming bacteria can change dramatically. In contrast to their approximately straight-line locomotion in bulk  fluids, non-tumbling flagellated bacteria typically follow circular trajectories near surfaces, often for an extended period of time~\cite{Frymier_EtAl_1995}. Furthermore, several wild-type peritrichous bacterial strains have been found to exhibit longer run times and smaller mean tumbling angles at the surface compared to their bulk run and tumbles~\cite{Molaei_EtAl_2014}. Exploiting  cell-surface interactions, recent studies demonstrated  that bacteria can be concentrated by funnel walls~\cite{Galajda_EtAl_2007}, drive asymmetric microgears \cite{DiLeonardo_EtAl_2010,Sokolov19012010}, and self-organize into collective vortices~\cite{Wioland_EtAl_2016, Nishiguchi_EtAl_2018}. A well-known consequence of bacteria-surface interactions is the accumulation of cells near solid surfaces:  Local concentration values for both non-tumbling and tumbling strains near a flat surface can exceed the corresponding bulk concentrations by a factor of 5 or more \cite{Berke_EtAl_2008, LiTang_2009, Li_EtAl_2011, Molaei_EtAl_2014}. Such accumulation increases the possibility cell-surface attachment, facilitating undesirable secondary effects like biofouling and biofilm formation~\cite{vanLoosdrecht_EtAl_1990, Conrad_2012}. Surface-attached microbial communities~\cite{Beroz:2018aa,2019HaEtAl_NatPhys_AOP} cause  widespread problems to a broad range of industrial equipment and infrastructure, such as food processing facilities \cite{KumarAnand_1998, ChmielewskiFrank_2003}, ships and pipes \cite{Flemming_2002}, and surgical equipment and medical implants \cite{Donlan_2001, SrivastavaBhargava_2016,Drescher4345}. In the medical context, these surface-attached microbial colonies are especially harmful because they can lead to persistent infection \cite{Costerton_EtAl_1999}. 
\par 
Over the past two decades, much progress has been made in designing antifouling surfaces based on chemical surface modification~\cite{Neu_1996, RennerWeibel_2011, Banerjee_EtAl_2011}. Common surface treatments include released-based coatings in which a biocidal agent (e.g. silver ions, antibiotics, or quaternary ammonium compounds) is released into the environment, hydrophilic polymer coatings, and self-assembled monolayers. However, the antifouling properties of these surface treatments are often temporary because of the depletion of the biocidal substance within the coating, or the masking of the coating's chemical functionality by the absorption of biomolecules from the surrounding environment \cite{Neu_1996, RennerWeibel_2011, Banerjee_EtAl_2011}. Further, chemical surface treatments can leach into and have toxic effects on the local ecosystem and have led to the rise of antibiotic- and silver-resistant bacterial strains \cite{Banerjee_EtAl_2011, RennerWeibel_2011}. Thus, chemical surface modifications alone are unlikely to provide long-term solutions to antifouling problem. An interesting alternative approach, inspired by the \textit{Nepenthes} pitcher plant, utilizes a lubricant-infused coating that results in a slippery surface \cite{Wong_EtAl_2011, Epstein_EtAl_2012}. Another nontoxic, persistent solution may be the manipulation of the surface topology to deter bacterial adhesion. Important previous studies of bacterial adhesion on various two-dimensional (2D) polydimethylsiloxane (PDMS) patterned surfaces have explored nanoscale tall spatially organized designs \cite{Perera-Costa_EtAl_2014},  shark skin inspired micrometer high diamond pattern \cite{Carman_EtAl_2006}, and nested hierarchically wrinkled surface topography with length scales spanning from tens of nanometers to a fraction of a millimeter \cite{Efimenko_EtAl_2009}. Yet, many aspects of the interplay between complex surface geometries and cell accumulation are not yet well understood.

\begin{figure*}[t]
\includegraphics[width=1\textwidth]{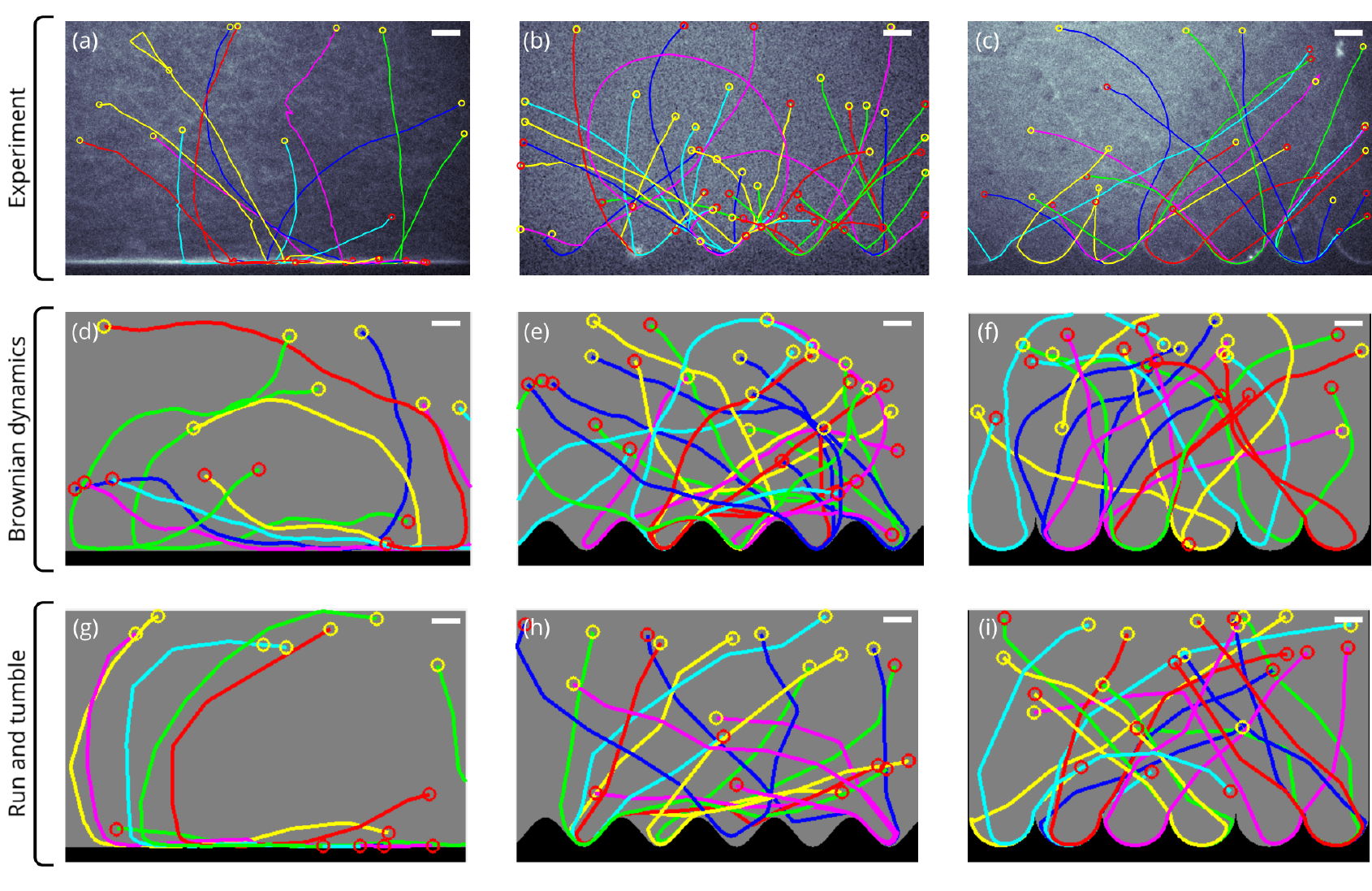}
\caption{\label{fig:traj} Typical trajectories of swimming cells for flat, sinusoidal, and semicircle surface geometries as observed in experiment and simulations.
The start and end of each trajectory are indicated by the yellow and red circle, respectively. Each trajectory is 10~s long. Bacteria align with the surface in the flat geometry leading to significant surface accumulation for the experiment and simulation models. The sinusoidal ($A = 5.25$ $\mu$m, $\lambda = 28$ $\mu$m) and concave semicircle ($R = 12$ $\mu$m) surface geometries redirect the bacteria away from the surface in the experiments and simulations. Scale bars 10 $\mu$m.
}
\end{figure*}

\par 
To contribute to a more detailed understanding, we investigate here in experiments and simulations  the effects of locally varying boundary curvature on the scattering and accumulation dynamics of swimming \textit{Escherichia coli} bacteria in quasi-2D microfluidic chambers (Fig.~\ref{fig:traj}). To explore the effects of partially convex and concave boundary geometries on the spatial cell distributions, we complement our experiments with simulations of a 2D particle-based simulations for both Brownian Dynamics (BD) and Run and Tumble (RT) dynamics. Our analysis confirms that a minimal steric interaction model~\cite{LiTang_2009, Li_EtAl_2011} suffices to account for main aspects of the experimental data. Both experimentally observed and simulated cell trajectories illustrate that the non-convex boundary features redirect the bacteria away from the surfaces (Fig.~\ref{fig:traj}). Throughout, data from experiments and simulations are analyzed using the same algorithms to compare the observed and predicted  surface accumulation (Fig.~\ref{fig:rawDataPoints}). Scanning a range of geometric surface parameters, we are able to determine an optimal curvature that minimizes the bacterial accumulation for a sinusoidal boundary geometry (Fig.~\ref{fig:sineHeatmap}). Furthermore, we find that periodic boundaries with a strictly concave base geometry can decrease the average cell accumulation near the boundary by more than 50\% relative to a flat surface (Fig.~\ref{fig:accHistFlatSineUUU}).

\section{Experiments}
We produced microfluidic chambers (4~mm long, 2~mm wide, 3-4~$\mu$m thick) by standard soft lithography technique from PDMS (Dow Corning). For the sinusoidal geometries (Fig.~\ref{fig:traj}b), the top and bottom boundaries of each chamber are designed as $\pm A \sin ( 2 \pi x / \lambda)$, and we investigated 20 different parameter combinations with amplitudes $A = [1.75, 3.5, 5.25, 7]$~$\mu$m and wavelengths $\lambda = [21, 28, 35, 42, 49]$ $\mu$m. The boundaries for the concave semicircle geometry (Fig.~\ref{fig:traj}c) were designed with radius $R = 12$ $\mu$m. To ensure the cell dynamics  and statistics are not biased by reflections from the opposing boundary, we chose a large boundary separation distance of $2$~mm. After a 40~s exposure to oxygen plasma (Harrick Plasma, PDC-002) the PDMS chambers were bonded to the glass coverslips initially cleaned in hydrogen peroxide. 
\par
Non-chemotactic \textit{E. coli} cells (strain HCB1733, provided by Howard C. Berg) carrying pYFP plasmid (Clontech, BD Biosciences) were streaked on 1.5\% agar plates containing Tryptone broth (TB, Sigma) and 100 $\mu$g/mL ampicillin. A single-colony isolate from an overnight plate was inoculated in 10 mL of the medium containing 10 ml TB, ampicillin and 0.1 mM isopropyl $\beta$-D-1-thiogalactopyranoside (IPTG, Sigma), which was then grown for 12~h on a rotary shaker (200 rpm) at $34^\circ$C. This culture was further diluted at 1:100 in the fresh medium and grown for further 4~h. The resulting culture was washed in the fresh medium with an addition of 0.1\% bovine serum albumin to prevent bacterial adhesion. The resulting bacterial suspension (approximately $10^8$ cells/mL) was loaded into the microfluidic chambers. The device inlets were then sealed with unpolymerised PDMS to avoid background fluid flow. The bacteria motion was measured using a Nikon TE2000U inverted microscope with a 40x oil immersion objective (NA 1.3) at 10 frames-per-second (Evolve Delta, Photometrics) or LSM 510 Zeiss Axiovert 200 M at 3 frames-per-second (fps). Single-cell trajectory data were reconstructed using a custom Matlab particle tracking script.

\section{Simulations}
To test whether steric surface collisions can account for the experimentally observed cell distributions, we performed 2D particle-based simulations. Focusing on minimal models, we neglect hydrodynamic effects~\cite{LiTang_2009, Li_EtAl_2011} because the small chamber thickness in our experiments strongly suppresses hydrodynamic flows. Similarly, steric cell-cell interactions can be ignored as we consider dilute bacterial suspensions throughout. The bacteria are modeled as non-interacting ellipsoids of half-length $\ell$ and half-width $r$, described by their position $\Tensor{x}(t)$ and orientation $\UnitVec{n}(t)$. Cells are assumed to move at a constant self-propulsion speed $v$ in the direction of their orientation $\UnitVec{n}$. An effective steric boundary potential $U$ is used to encode bacterial surface interactions across various geometries. Bacteria in the experiments display occasional stochastic reorientation as they swim \cite{Drescher_EtAl_2011}. To account for this,  we perform and compare simulations for both BD and RT reorientation. In the BD model, bacteria are reoriented through Gaussian rotational noise. In the RT model, a cell moves deterministically for a fixed period of time (run stage) before undergoing a stochastic reorientation event (tumble stage).

\subsection{Brownian Dynamics (BD)}
Denoting the $\mathtt{d}$-dimensional unit matrix by $\Tensor{I}$, the over-damped Langevin equations for a single bacterium with position $\Tensor{x}(t)$ and orientation $\UnitVec{n}(t)$ in the BD model are
\begin{subequations} 
\label{eqn:BrownianDyn}
\begin{align}
\mathrm{d}\Tensor{x} &= ( v \UnitVec{n} - \Tensor{\Gamma}^{-1}\nabla_\Tensor{x}U ) \mathrm{d}t \label{eqn:BrownianDyn_x} \\
\mathrm{d}\UnitVec{n} &= ( \Tensor{I} - \UnitVec{n} \UnitVec{n}^\top ) \left( (1 - \mathtt{d}) D_R \UnitVec{n} - \Tensor{\Omega} \nabla_\UnitVec{n}U \right) \mathrm{d}t \nonumber \\
	&\quad + \sqrt{2 D_R} ( \Tensor{I} - \UnitVec{n} \UnitVec{n}^\top ) \cdot \mathrm{d} \Tensor{Z} \label{eqn:BrownianDyn_n}
\end{align}
\end{subequations}
Here, $v$ is the self-swimming speed, $D_R$ the rotational diffusion coefficient, and $\Tensor{Z}$ is a $\mathtt{d}$-dimensional Gaussian random variable of zero mean and variance $\mathrm{d}t$. The $(1 - \mathtt{d}) D_R \UnitVec{n}$ term is required to ensure that Eq. \eqref{eqn:BrownianDyn_n} preserves the unit length of $\UnitVec{n}$,  i.e. $\mathrm{d}|\UnitVec{n}|^2 = 0$. The boundary potential $U$ used for the cell-surface interactions will be described in detail below. The friction tensor
\begin{equation}
\Tensor{\Gamma} = \gamma_0 \left[ \gamma_\parallel (\UnitVec{n} \UnitVec{n}^\top) + \gamma_\perp ( \Tensor{I} - \UnitVec{n} \UnitVec{n}^\top ) \right]
\end{equation}
accounts for the fact that the bacteria experience more drag when moving perpendicular to their orientation. Rotational drag is approximated as isotropic,
\begin{equation}
\Tensor{\Omega} = \frac{1}{\omega_0} \frac{1}{\gamma_R} \Tensor{I}.
\end{equation}
$\gamma_0$ and $\omega_0 = {k_B T}/{D_R}$ are the Stokesian translational and rotational friction coefficients, respectively. $k_B$ is the Boltzmann constant, and $T$ is the temperature.
$\gamma_\parallel$, $\gamma_\perp$, and $\gamma_R$ are dimensionless geometric parameters characterizing the longitudinal, transverse, and rotational friction parameters of elongated particles that depend only on the aspect ratio $a = \ell / d$. We use the expressions  given in \cite{Tirado_EtAl_1984} for rod-like macromolecules
\begin{subequations}
\label{eqn:nonDimDragFactors}
\begin{align}
\frac{2 \pi a}{\gamma_\parallel} &= \ln a - 0.207 + \frac{0.980}{a} - \frac{0.133}{a^2} \\
\frac{4 \pi a}{\gamma_\perp} &= \ln a + 0.839 + \frac{0.185}{a} + \frac{0.233}{a^2} \\
\frac{\pi a^2}{3 \gamma_R} &= \ln a - 0.662 + \frac{0.917}{a} - \frac{0.050}{a^2}
\end{align}
\end{subequations}

Adopting cell length $\ell$ and $\tau = \ell / v$ as characteristic length and time scales and defining the following P{\'{e}}clet numbers $P_T \equiv v \ell \gamma_0 / k_B T$ and $P_R \equiv v / D_R \ell$, we can recast Eq. \eqref{eqn:BrownianDyn} in nondimensional form. Denoting dimensionless quantities with a superscript $^*$, we have
\begin{widetext}
\begin{subequations} \label{eqn:nonDimBrownianDyn}
\begin{align}
\mathrm{d}\Tensor{x}^* &= \left( \UnitVec{n}  - \frac{\epsilon}{k_B T} \frac{1}{P_T}  \left[ \frac{1}{\gamma_\parallel}(\UnitVec{n} \UnitVec{n}^\top) + \frac{1}{\gamma_\perp}( \Tensor{I} - \UnitVec{n} \UnitVec{n}^\top ) \right] \nabla_{\Tensor{x}^*} U^* \right) \mathrm{d}t^* \\
\mathrm{d}\UnitVec{n} &= ( \Tensor{I} - \UnitVec{n} \UnitVec{n}^\top ) \left( (1 - \mathtt{d}) \frac{1}{P_R} \UnitVec{n} - \frac{\epsilon}{k_B T} \frac{1}{P_R} \frac{1}{\gamma_R} \nabla_\UnitVec{n} U^* \right) \mathrm{d}t^* + \sqrt{\frac{2}{P_R} \mathrm{d}t^*} ( \Tensor{I} - \UnitVec{n} \UnitVec{n}^\top ) \cdot \mathrm{d} \Tensor{Z}
\end{align}
\end{subequations}
where $\epsilon$ characterizes the strength of the bacteria-boundary potential interaction.

\begin{figure*}[t]
\includegraphics[width=1\textwidth]{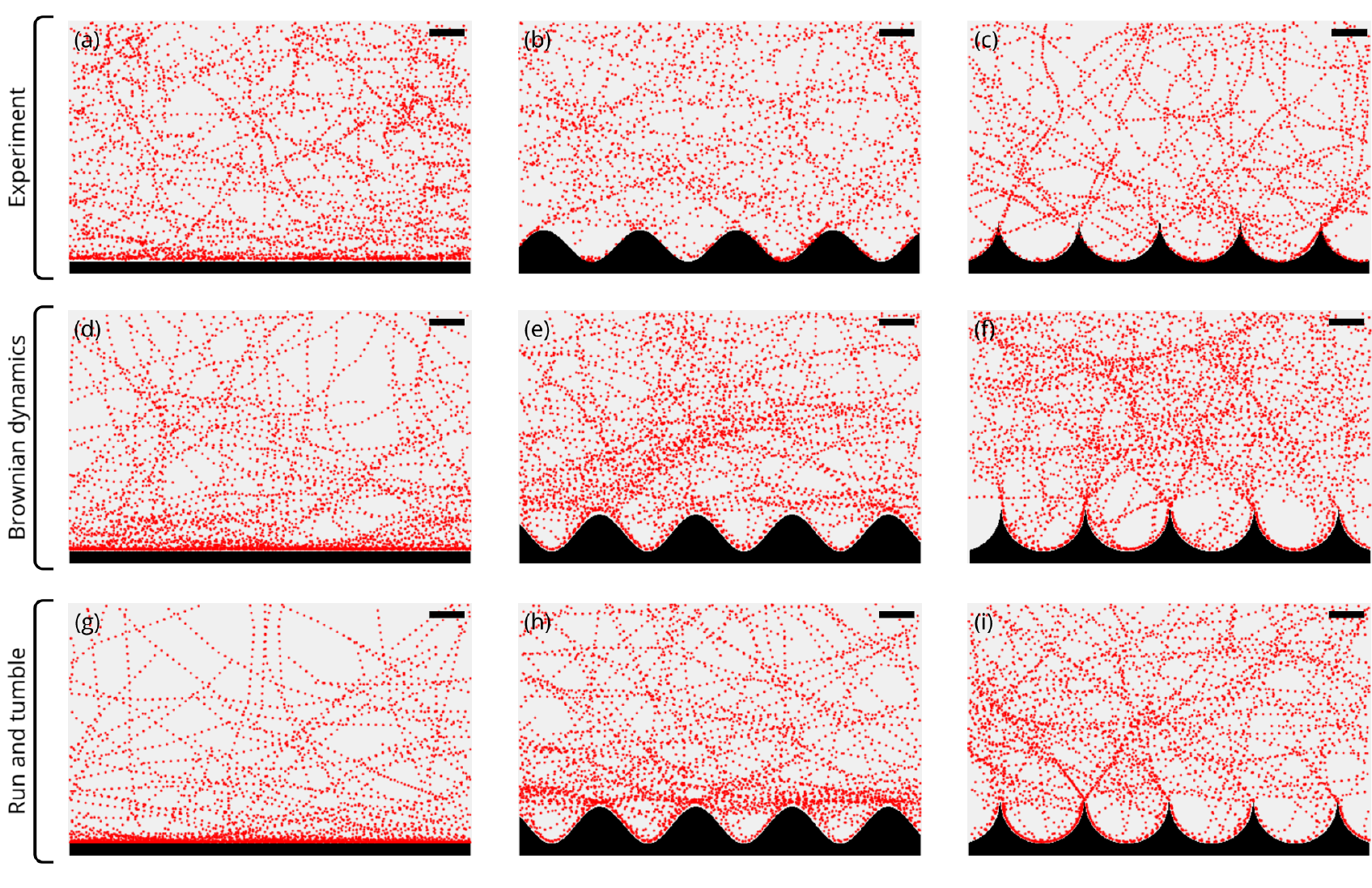}
\caption{\label{fig:rawDataPoints} Segmented raw data for the experiment and simulations, used in the statistical analysis.
The segmented trajectories are acquired at 10 fps. The experimental raw data exhibit higher curvature than the simulation raw data likely due to hydrodynamic effects, which are not accounted for in the simulations. Scale bar 10 $\mu$m.
}
\end{figure*}

\subsection{Run and Tumble (RT)}
We aim to compare the BD model with a corresponding RT model.  During the run stage of the RT model, which lasts a duration $\tau_\text{run}$,  the deterministic motion of a cell is governed by Eq. \eqref{eqn:BrownianDyn} with $D_R = 0$. Rescaling to a dimensionless form  using  the same characteristic length and time scales as before, the run motion is described by
\begin{subequations} \label{eqn:nonDimRunTumbleDyn}
\begin{align}
\mathrm{d}\Tensor{x}^* &= \left( \UnitVec{n}  - \frac{\epsilon}{k_B T} \frac{1}{P_T}  \left[ \frac{1}{\gamma_\parallel}(\UnitVec{n} \UnitVec{n}^\top) + \frac{1}{\gamma_\perp}( \Tensor{I} - \UnitVec{n} \UnitVec{n}^\top ) \right] \nabla_{\Tensor{x}^*} U^* \right) \mathrm{d}t^* \\
\mathrm{d}\UnitVec{n} &= ( \Tensor{I} - \UnitVec{n} \UnitVec{n}^\top ) \left( - \frac{\epsilon}{k_B T} \frac{1}{P_R} \frac{1}{\gamma_R} \nabla_\UnitVec{n} U^* \right) \mathrm{d}t^*
\end{align}
\end{subequations}
\end{widetext}
where the rotational P{\'{e}}clet number due to tumbling is now determined as follows: At the end of $\tau_\text{run}$, the bacterium undergoes a tumbling event. Let $\theta$ be the angle between the previous orientation and new orientation after a tumble. $\theta$ is drawn from a von Mises (vM) distribution with the mean angle equal to the original bacterial orientation and concentration parameter $\kappa$. To relate $\kappa$ to experimental values, we note that for weakly tumbling cells $\kappa \gg 1$. In this case,  the mean squared angular change per tumble is $\langle \theta^2\rangle=\tilde{D}_R  \tau_\text{run}\simeq 1/\kappa$, yielding the effective rotational P{\'{e}}clet number $P_R=v/(\tilde{D}_R\ell)$.

\subsection{Boundary interactions}
The boundary potential $U$ prevents the bacteria from penetrating the boundary and forces them to align parallel with the local surface tangent. This is achieved by penalizing the overlap between the bacteria and the surface exponentially
\begin{equation}
U = 
	\begin{cases}
	0 & \quad \text{if } z \leq 0 \quad \text{`no contact'}\\
	\epsilon e^{z / \sigma} & \quad \text{if } z > 0 \quad \text{`contact'}\\
	\end{cases}
\end{equation}
where $\epsilon$ is the strength parameter for the bacteria-boundary interaction and $\sigma$ is a length scale parameter of the order of the bacterial width. The overlap coordinate $z$ is defined as 
\begin{equation}
z = \ell | \UnitVec{n} \cdot \UnitVec{N}(\Tensor{x}) | + r - \UnitVec{N}(\Tensor{x}) \cdot \left( \Tensor{x} - \Tensor{S} (\Tensor{x}) \right)
\end{equation}
$\Tensor{S} (\Tensor{x})$ is the point on the surface that is closest to the bacterium's position $\Tensor{x}$, and $\UnitVec{N}(\Tensor{x})$ is the surface normal vector at $\Tensor{S} (\Tensor{x})$. Recall, $\ell$ and $r$ are the bacterium's half-length and half-width, respectively. The first term in $z$ is the projected half-length in the direction of the surface normal, and the last term is the signed distance of the bacterium's center from the surface. Explicit expressions for the derivatives of the boundary potential with respect to $\Tensor{x}$ and $\UnitVec{n}$ are given in Appendix \ref{sec:boundaryPotentialDerivatives}.

\subsection{Implementation}
We consider mirror-symmetric confinements parallel to the $y = 0$ line defined by surfaces $s_{y \pm} = f_\pm(s_x)$ with $f_-(s_x) = -f_+(s_x)$ where $\Tensor{S} = (s_x, s_y)$ denotes a point on the surface. The distance, $d$, of a bacterium at a position $\Tensor{x} = (x, y)$ from a surface $f$ is given by the function 
\begin{equation}
d(s_x) = \frac{1}{2} \left[ (x - s_x)^2 + (y - f(s_x))^2 \right]
\end{equation}
where the numerical prefactor $1/2$ was chosen for convenience. To find the point on the surface closest to the bacterium, we solve 
\begin{equation}\label{eqn:ClosestSurfacePoint}
\frac{\partial d}{\partial s_x} = s_x - x + (s_y - y) \frac{\partial s_y}{\partial s_x} = 0
\end{equation}
numerically with the bisection method and use the second derivative to confirm that the surface point found results in a minimum distance. The boundary surface equations for the flat, sinusoidal, and semicircle surfaces used in the simulations are
\begin{equation}
s_{y \pm} = \pm C
\end{equation}
\begin{equation}
s_{y \pm} = \pm A \sin \left( \frac{2 \pi}{\lambda} s_x \right) \pm C
\end{equation}
\begin{equation}
\label{eqn:semicircleSurface}
s_{y \pm} = \pm \sqrt{R^2 - \left\{ \frac{2R}{\pi} \cos^{-1} \left[ \cos \left( \frac{\pi}{2R} s_x \right) \right] - R \right\}^2} \pm C
\end{equation}
$C = 1000$ $\mu$m is the displacement from the $y=0$ line for all geometries. Because there is a discontinuity in the derivative of the semicircle geometry at the peaks ($s_x = 2Rn \text{ for } n = 0, 1, 2, ...$), we neglected the boundary potential for a region of scale $\sim r$ at the peaks for bacteria that are not vertical and treated peaks as a flat boundary for bacteria that are vertical to prevent the cells from penetrating the surface.
\par
A parallel individual-based code was developed to perform the simulations on a graphics processing unit (GPU). At each time step, the new positions and orientations of the bacteria are obtained from solving the dimensionless over-damped translation and orientation equations for the BD and RT models, Eqs. \eqref{eqn:nonDimBrownianDyn} and \eqref{eqn:nonDimRunTumbleDyn}. The numerical integration is performed using the Euler scheme, and  $\UnitVec{n}$ is renormalized at each time step to correct for integration errors. Cells are initially loaded uniformly within the computational domain with random orientations and with random start times for the run time for the RT model. Periodic boundary conditions were applied in the $x$-direction. Measurements were taken after the simulations had relaxed to a statistical  steady-state with constant  $\langle y^2\rangle$.

\begin{table*}[t]
\centering
\setlength\doublerulesep{1pt}
\begin{tabular}{l | c | c | p{0.5\linewidth}}
			& BD  					& RT						& Description 		\\ \hhline{=|=|=|=}
$\ell$ 		& 3.5 $\mu$m 			& 3.5 $\mu$m   			& Bacteria half-length 	\\				
$r$ 		& 0.5 $\mu$m			& 0.5 $\mu$m   			& Bacteria half-width					\\
$v$			& 20 $\mu$m/s			& 20 $\mu$m/s				& Self-propulsion speed				 \\
$\epsilon$	& 175  $k_B T$			& 1500  $k_B T$  			& Boundary potential strength parameter \\
$\sigma$	& 0.5 $\mu$m			& 0.5 $\mu$m				& Boundary potential scale parameter	\\
$D_R,\tilde{D}_R$		& 0.08 $\text{rad}^2/s$ & 0.1 $\text{rad}^2/s$		& Rotational diffusion coefficient\\
$P_T$		& 0.0014				& 0.0014					& Translational  P{\'{e}}clet number \\
$\tau_\text{run}$ & --				& 1 s						& Run time								\\
$\kappa$ 	& --					& 10					    & Concentration parameter for von Mises-Fisher distribution	
\end{tabular}
\caption{\label{tab:simParams} Summary of simulation parameters}
\end{table*}

\subsection{Parameters}
We model the bacteria as 1 $\mu$m in width and 7 $\mu$m in length, accounting for part of the flagellum in addition to cell body length. It is known that \textit{E. coli} move at a speed of approximately 20 $\mu$m/s \cite{Drescher_EtAl_2011}, and the run time is typically 1 s \cite{BergBrown_1972, Turner_EtAl_2000}. Simulation scans were performed to find $\epsilon$ and $D_R$ that resulted in surface accumulations that best matched with the experiments for the sinusoidal surface. For the BD model, we found $\epsilon = 175$ $k_B T$ and $D_R = 0.08$ $\text{rad}^2/s$, and  for the RT model $\epsilon = 1500$ $k_B T$ and $D_R = 0.1$ $\text{rad}^2/s$. 
The fitted near-surface $D_R$ values for both models are of the same order of magnitude as the measured bulk $D_R = 0.057$ $\text{rad}^2/s$ for non-tumbling \textit{E. coli} \cite{Drescher_EtAl_2011}.  For both models $\epsilon \gg k_B T$, indicating that the boundary potential is highly repulsive. The large $\epsilon$ is required to prevent the bacteria from penetrating the boundary as the models do not account for the reduction in swimming speed as the cells approach the surface. The fitted $\epsilon$ and $D_R$ values obtained for the sinusoidal surface were also used for simulations of the flat and semicircle surfaces. A summary of all relevant simulation parameters  is given in Table~\ref{tab:simParams}.
\section{Results}
\begin{figure*}[t]
\includegraphics[width=1\textwidth]{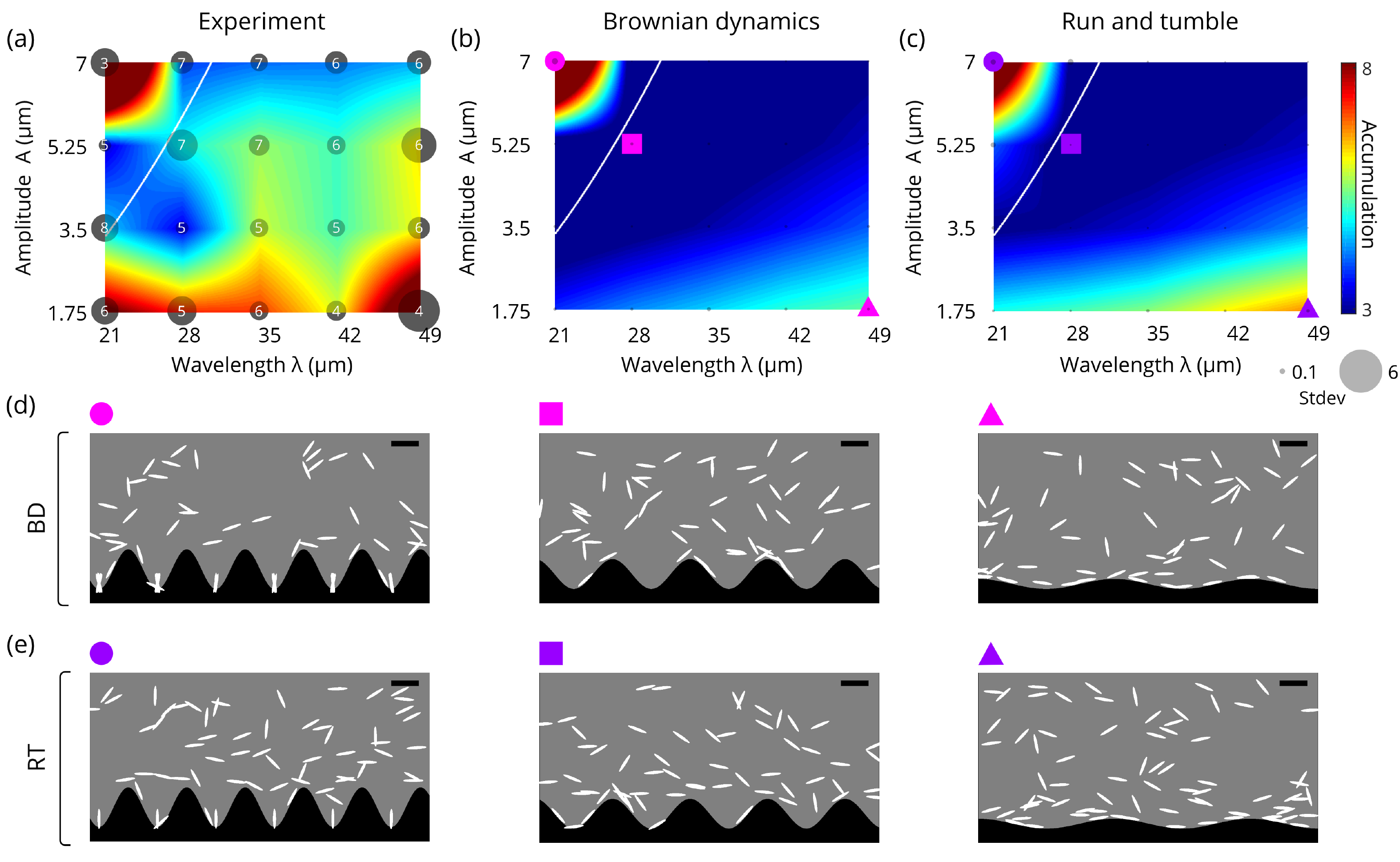}
\caption{\label{fig:sineHeatmap} Mean bacteria surface accumulation for the sinusoidal surface over a range of amplitudes $A$ and wavelengths $\lambda$.
Accumulation at the surface is measured by comparing the number of bacteria within $5$ $\mu$m from the surface to the number of bacteria in the same area $50$ $\mu$m away from the surface (Fig.~\ref{fig:accHistFlatSineUUU}). (a) The location of the circles indicate the 20 combinations $(A,\lambda)$ of the scan, and the size of the circle represents the standard deviation of each point. The white numbers indicate the number of experiments per point. (b,c) 3 simulations were performed for the same pairs $(A,\lambda)$ as in the experiments and bilinearly interpolated. The BD and RT simulations agree qualitatively with experiment, revealing an optimum max curvature that reduces accumulation. The set of parameters corresponding to the optimum curvature $\kappa^*$ is delineated by the white curve $A = (\kappa^* / 4\pi^2) \lambda^2$ where $\kappa^* = 0.31$ $\mu$m$^{-1}$. Typical images for the BD and RT simulations are shown in (d) and (e), respectively, for $A = 7$~$\mu$m, $\lambda = 21$ $\mu$m (circle), $A = 5.25$~$\mu$m, $\lambda = 28$ $\mu$m (square), and $A = 1.75$~$\mu$m, $\lambda = 49$ $\mu$m (triangle). Scale bars 10 $\mu$m.
}
\end{figure*}

For the experiments and simulations, the cell trajectories (Fig. \ref{fig:traj}) of the flat, sinusoidal, and semicircle geometries are segmented (Fig. \ref{fig:rawDataPoints}) to quantify the bacterial accumulation. To identify optimal sinusoidal surfaces for the reduction of bacterial surface accumulation, we performed a scan over a range of amplitudes and wavelengths (Fig. \ref{fig:sineHeatmap}). The cell distributions of the optimal sinusoidal and semicircle surfaces are quantified and compared, with the semicircle geometry proving to be the most efficient at reducing bacterial accumulation (Fig. \ref{fig:accHistFlatSineUUU}).

\subsection{Tracking data}
The simulated particle trajectories agree well with the experimental cell trajectories (Fig. \ref{fig:traj}). In the flat geometry, bacteria collide and align with the surface~\cite{LiTang_2009, Li_EtAl_2011}, contributing to surface accumulation in the experiments and simulations. Comparing Fig. \hyperref[fig:traj]{\ref*{fig:traj}(d) and \ref*{fig:traj}(g)}, we note bacterial residence time at the surface appears shorter in the BD model than the RT model due to the orientation noise. Because of the non-convex features present in both the sinusoidal and semicircle geometries, the bacteria are redirected away from the surface in both the experiment and simulations, leading to a reduction in surface accumulation. The segmented raw data is normalized to the same frame rate for both the experiment and simulations (Fig. \ref{fig:rawDataPoints}). This allows us to examine experimental and simulation data using the same analysis algorithms. The experimental raw data Fig. \hyperref[fig:rawDataPoints]{\ref*{fig:rawDataPoints}(a) - \ref*{fig:rawDataPoints}(c)} exhibit higher curvature than the simulation raw data, likely caused by hydrodynamic effects from the channel walls, which are not accounted for in the simulations.

\subsection{Optimal sinusoidal boundaries}
\begin{figure*}[t]
\includegraphics[width=0.8\textwidth]{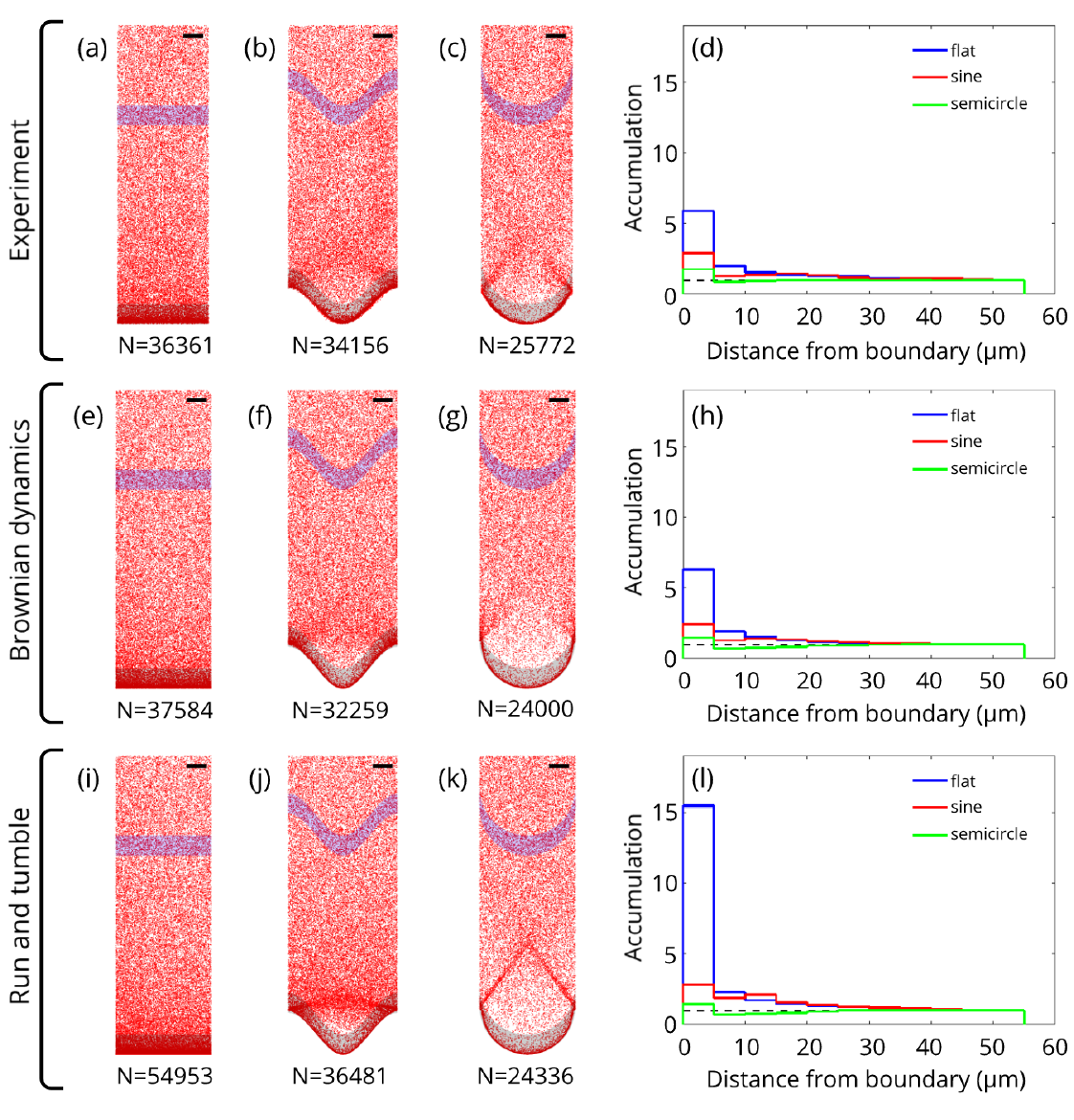}
\caption{\label{fig:accHistFlatSineUUU} Sampled raw data and accumulation histograms for the flat, sinusoidal ($A = 5.25$ $\mu$m, $\lambda = 28$ $\mu$m), and concave semicircle ($R = 12$ $\mu$m) surface geometries for the experiment and simulations.
To visualize the spatial cell distributions, the raw data, acquired at 10 fps for 5 min, were projected onto a single wavelength (the flat surface is assumed to have the same wavelength as the semicircle surface) and sampled such that the bulk density is the same in all cases (a) - (c), (e) - (g), and (i) - (k). Both the experiments and simulations qualitatively show a depletion zone above the boundary for the sinusoidal and semicircle geometries. Due to the differences in the surface accumulation, the total cell numbers differ for the three geometries. The accumulation histograms (d), (h), and (l) quantify this effect, with accumulation defined is the ratio of the number of bacteria in each surface bin area (grey region for first bin) to the number of bacteria in a congruent area $50$ $\mu$m away from the surface (blue region). The results are independent of the shape of the bulk reference area (see Fig. \ref{fig:accBarFlatSineUUU}). Histograms (d), (h) and (l) were computed from  20 independently subsamples of the raw data. The bulk accumulation value 1 is indicated by the dashed black line. The accumulation histograms show that the concave semicircle geometry is the most efficient at suppressing accumulation in the experiment and simulations. Bin width 5 $\mu$m. Scale bars 5 $\mu$m.
}
\end{figure*}

To determine optimal sinusoidal boundary geometries, we perform a parameter scan over a range of amplitudes $A$ and  wavelengths $\lambda$, measuring cell accumulation at the surface in each case [Fig. \hyperref[fig:sineHeatmap]{\ref*{fig:sineHeatmap}(a)]}.
The bacterial surface concentration is determined from the number of cells between the surface boundary and the boundary contour shifted 5 microns away from the boundary (Fig.~\ref{fig:accHistFlatSineUUU}). The bulk concentration reflects the number of bacteria in a congruent area  50 $\mu$m away from the boundary. Accumulation is quantified as the ratio of the surface concentration over the bulk concentration. Figures \hyperref[fig:sineHeatmap]{\ref*{fig:sineHeatmap}(a) - \ref*{fig:sineHeatmap}(c)} illustrate the resulting mean surface accumulation of the scan. The location and size of the grey circles in Figs. \hyperref[fig:sineHeatmap]{\ref*{fig:sineHeatmap}(a) - \ref*{fig:sineHeatmap}(c)} designate the 20 combinations $(A, \lambda)$ and the standard deviation, respectively. In Fig. \hyperref[fig:sineHeatmap]{\ref*{fig:sineHeatmap}(a)}, the white numbers indicate number of experiments performed per point. Simulations were performed for the same 20 combinations $(A,\lambda)$ as in  the experiment;  for Figs. \hyperref[fig:sineHeatmap]{\ref*{fig:sineHeatmap}(b) - \ref*{fig:sineHeatmap}(c)},  are based on 3 simulations per point.
\par 
As evidenced by the mean surface accumulation, both the BD and RT models agree qualitatively with the experiment. Typical still images from the simulations are shown in Figs. \hyperref[fig:sineHeatmap]{\ref*{fig:sineHeatmap}(d) - \ref*{fig:sineHeatmap}(e)}  for $A = 7$~$\mu$m, $\lambda = 21$~$\mu$m (circle), $A = 5.25$~$\mu$m, $\lambda = 28$~$\mu$m (square), and $A = 1.75$~$\mu$m, $\lambda = 49$~$\mu$m (triangle). Due to the steep curvature of the sinusoidal boundary at $A = 7$~$\mu$m, $\lambda = 21$~$\mu$m, the cells become trapped in the surface pockets, leading to increased accumulation. The BD and especially the RT model can also capture the high accumulation at $A = 1.75$~$\mu$m, $\lambda = 49$~$\mu$m. Here, the surface is nearly flat and does not deflect the bacteria away from the surface, resulting in high accumulation. Quantitative differences between the experiment and simulations can likely  be attributed to hydrodynamic effects.
The low-accumulation region in both the experiment and simulations suggests that there exists an optimal curvature for suppressing bacterial accumulation. Characterizing this effect in terms of the maximal local curvature $\kappa^*$ of the sine wave, we find the relation
\begin{equation} \label{eqn: sineOptCurv}
A = (\kappa^* / 4\pi^2) \lambda^2
\end{equation}
After smoothing the experimental values with bilinear interpolation, we fit all the points that are within 15\% of the minimum accumulation to Eq. \eqref{eqn: sineOptCurv} and find the optimal maximal curvature $k^* = 0.31$ $\mu$m$^{-1}$. This is plotted as the white curve in Figs. \hyperref[fig:sineHeatmap]{\ref*{fig:sineHeatmap}(a) - \ref*{fig:sineHeatmap}(c)}. 

\subsection{Sinusoidal vs. concave geometries}
Previous work has shown that bacteria can be trapped by convex walls~\cite{Sipos_EtAl_2015}. This suggests that surface accumulation could be suppressed even further by replacing the sinusoidal boundaries with strictly concave structures. To test this hypothesis, we created the strictly non-convex semicircle geometry ($R = 12$ $\mu$m) seen in Fig. \hyperref[fig:traj]{\ref*{fig:traj}(c)}.
To compare this semicircle surface with the flat and the optimal sinusoidal ($A = 5.25$ $\mu$m, $\lambda = 28$ $\mu$m) surfaces, the segmented bacteria trajectories (acquired at 10 fps for 5 min) are projected onto one wavelength. We set the flat surface to have the same wavelength as the semicircle geometry. Because the bacterial concentration is different for the three surfaces, we normalize the raw data by using Bernoulli sampling to ensure the bulk density, defined as the density $50$ $\mu$m away from boundary, is the same in each case. Samples of the resulting normalized data in Fig. \ref{fig:accHistFlatSineUUU} illustrate the distribution of cells for the three surfaces for the experiment and simulations. The total cell numbers differ between the three geometries reflecting the differences in the surface entrapment. After contact with the sinusoidal and semicircle geometries, the bacteria leave the surface at a particular angle, as evidenced by the inward streaks in Figs.~\hyperref[fig:accHistFlatSineUUU]{\ref*{fig:accHistFlatSineUUU}(b) -  \ref*{fig:accHistFlatSineUUU}(c)}. This behavior is more clearly reproduced in the RT simulations than the BD simulations. Above the boundaries, we note the existence of depletion zones for the sinusoidal and, more prominently, the semicircle geometry. The fact that these depletion zones are also reproduced by our simulations suggests that they arise from the scattering dynamics and not by hydrodynamic effects.
\par
The accumulation histograms in Figs. \hyperref[fig:accHistFlatSineUUU]{\ref*{fig:accHistFlatSineUUU}(d), \ref*{fig:accHistFlatSineUUU}(h), and \ref*{fig:accHistFlatSineUUU}(l)} quantify and compare the bacterial distribution, where the solid line and shaded regions represent the mean and standard deviation of the accumulation ratio for 20 samples of the raw data. As before, accumulation is defined as the number of bacteria in the each bin area (grey region for the first bin near the surface) divided by the number of bacteria in the bulk area (blue region), which is $50$ $\mu$m away from the surface. Each bin is $5$ $\mu$m tall and follows the surface geometry. The results are independent of the shape taken for the bulk area (Fig. \ref{fig:accBarFlatSineUUU}). The black dashed line at height 1 indicates the bulk reference value. As evident from Fig. \hyperref[fig:accHistFlatSineUUU]{\ref*{fig:accHistFlatSineUUU}(d)}, both the flat geometry (blue line) and the partially convex sinusoidal geometry (red line) lead to cell accumulation above the bulk level within up to 30 $\mu$m from the surface, although this effect is substantially weaker for the sinusoidal geometry. By contrast, except very close to the surface, the distribution of cells for the semicircle geometry (green line) is at the bulk level. Close to the surface, the semicircle geometry decreases the average cell concentration by 70\% relative to a flat surface. Thus, the concave semicircle geometry is the most efficient at suppressing accumulation, in agreement with the predictions from the BD and RT models.
\par
Compared to the RT model, the cells in the BD model leave the surface more easily. This can be seen in the histogram curves for the flat geometries (blue lines), which show good agreement between the BD model and experiment, whereas the RT model overestimates the accumulation in the first bin  (Fig. \ref{fig:accBarFlatSineUUU}l). Yet, the RT model performs slightly better at replicating the trajectories of the cells after contact with curved surfaces than the BD model  (Fig. \ref{fig:accBarFlatSineUUU}c,g,k). Thus, bacterial reorientation in our experiments is likely a combination of BD and RT. While the BD and RT underestimate the accumulation for the sinusoidal geometry, they both agree well with experiment for the semicircle geometry, suggesting that near-field hydrodynamics could play a larger role in the bacterial surface entrapment for flat and convex geometries than for concave geometries.

\section{Conclusion}
Combining experiments and simulations, we studied the scattering and accumulation dynamics of swimming bacteria in the vicinity of curved periodic boundaries. Our results demonstrate that a concave boundary can reduce the average cell accumulation by more than 50\% relative to a flat surface. Despite the simplifying model assumptions, we found that simulations of a basic steric interaction model can account  for the experimental observations across the different geometries. In the future, it would be interesting to perform a similar analysis for 2D microtopographic surface designs.

\begin{acknowledgments}
This work was supported by the MIT OGE Chyn Duog Shiah Memorial Fellowship (R.M.), a James S. McDonnell Complex Systems Scholar Award  (J.D.), and a Royal Society Research Grant RG150072 (V.K.). We thank Howard C. Berg for kindly providing the \textit{E. coli} strains.
\end{acknowledgments}

%


\appendix
\onecolumngrid


\section{\label{sec:boundaryPotentialDerivatives}Boundary potential derivatives}
The translational and rotational gradients of $U$ are $
\partial_{x_i} U = \partial_{z} U \partial_{x_i} z
$ and
$
\partial_{\hat{n}_i} U = \partial_{z} U \partial_{\hat{n}_i} z
$
, where
$
\partial_{z} U = ({\epsilon}/{\sigma}) e^{z / \sigma}.
$
The translational gradient of $z$ is
\begin{equation} \label{eqn:TransGrad_z}
\partial_{x_i} z = \left( \ell \frac{\UnitVec{n} \cdot \UnitVec{N}(\Tensor{x})}{| \UnitVec{n} \cdot \UnitVec{N}(\Tensor{x}) |} \hat{n}_j - x_j + S_j \right) \partial_{x_i} \hat{N}_j + \hat{N}_j \partial_{x_i} S_j - \hat{N}_i
\end{equation}
If the surface is flat, both $\UnitVec{N}$ and $\Tensor{S}$ are constant and independent of the bacterium's position $\Tensor{x}$; thus, Eq. \eqref{eqn:TransGrad_z} simplifies to $\partial_{x_i} z  = -\hat{N}_i$. Hence, for flat surfaces, the translational force is in the direction of the surface normal. The rotational gradient of $z$ is
\begin{equation}
\partial_{\hat{n}_i} z = \ell \frac{\UnitVec{n} \cdot \UnitVec{N}(\Tensor{x})}{| \UnitVec{n} \cdot \UnitVec{N}(\Tensor{x}) |} \hat{N}_i
\end{equation}


\begin{figure*}[h]
\includegraphics[width=0.625\textwidth]{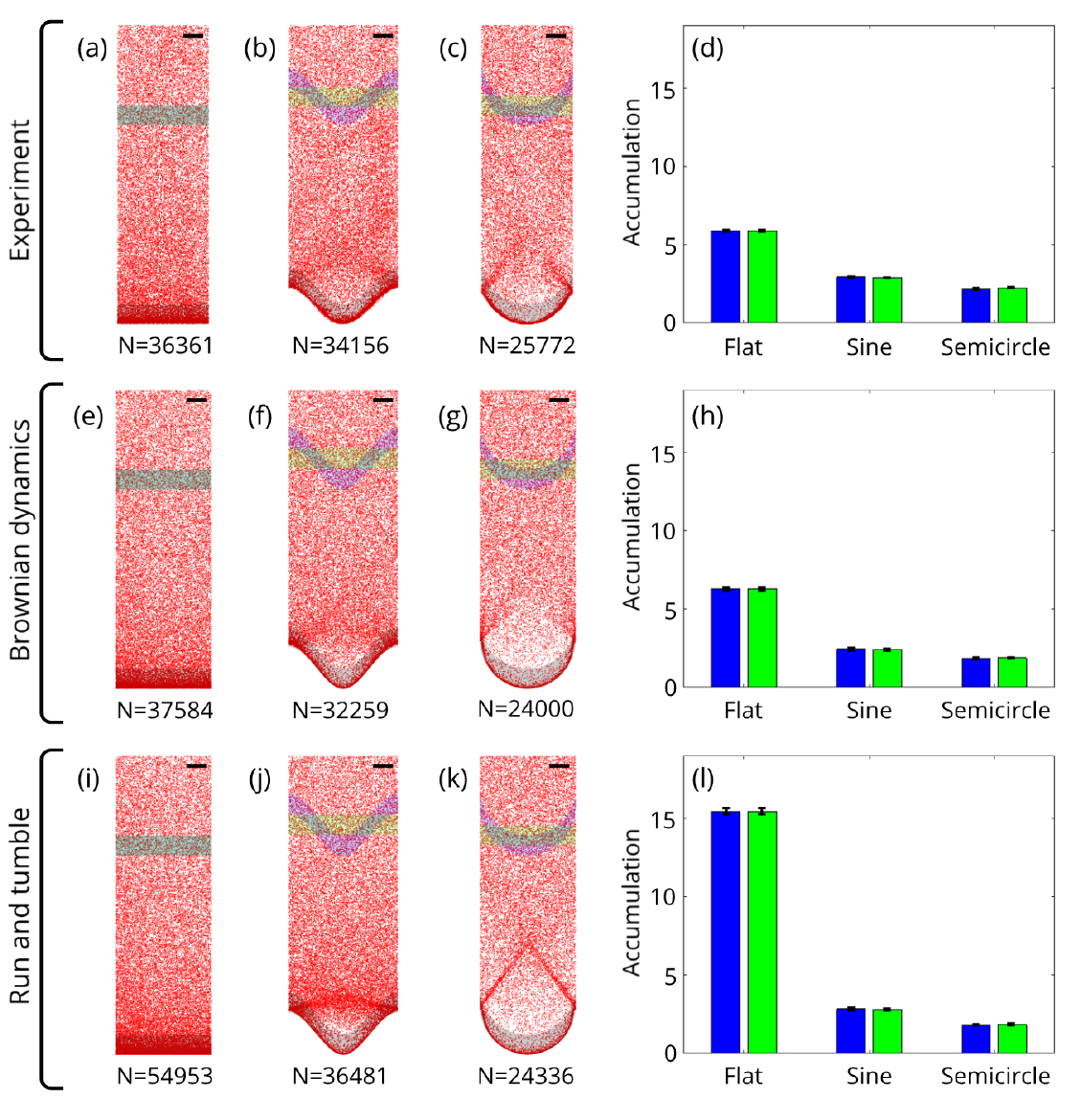}
\caption{\label{fig:accBarFlatSineUUU} Sampled raw data and surface accumulation bar graphs for the flat, sinusoidal ($A = 5.25$ $\mu$m, $\lambda = 28$ $\mu$m), and semicircle ($R = 12$ $\mu$m) surface geometries for the experiment and simulations.
To visualize the spatial cell distributions, the raw data, acquired at 10 fps for 5 min, were projected onto a single wavelength (the flat surface is assumed to have the same wavelength as the semicircle surface) and sampled such that the bulk density is the same in all cases (a) - (c), (e) - (g), and (i) - (k). Due to the differences in the surface accumulation, the total cell numbers differ for the three geometries.
Accumulation is defined as the ratio of the number of bacteria in each surface bin area (grey regions)  to the number of bacteria in an equally sized area $50$ $\mu$m away from the surface. Two shapes of equal area are considered for the bulk area: a shape which follows the surface (blue) and a rectangle (green). For each geometry in (d), (h), and (l), the blue and green bar show the mean surface accumulation calculated with the surface shape and rectangle as the bulk area, respectively, for 20 samples of the raw data.
The error bars represent the standard deviation. The blue and green bars are nearly equal for each case, demonstrating that the surface accumulation estimation is independent of the shape taken for the bulk reference area. Bin width 5 $\mu$m. Scale bars 5~$\mu$m.
}
\end{figure*}

\end{document}